# About the Problem of Utilization the Low-Potential Heat and Recent Perspective Developments

## Ivan V. Kazachkov [1,2]

[1] *Dept of Information technology and data analysis, Nizhyn Gogol state university,*
*Grafs'ka, 2, Ukraine, 16600, Tel: +380973191343;Email: ivan.kazachkov@energy.kth.se*
[2] *Dept of Energy Technology, Royal Institute of Technology, Sweden*

**Abstract**

Problem of utilization of the low-potential heat and its perspectives are considered and one example from the author's works is presented in detail. Description of the project and the results obtained are presented by a development of the Liquid Metal Magnetohydrodynamic (LMMHD) Gravitational Mini Power Plant (GMPP) for utilization of the low-potential heat from any available low-exergy sources, which are the huge sources of the wasted energy around the globe. Project is based on the experience by the LMMHD energy transformation with gravitational vapour/gas-lift driving principle. An example of the GMPP was developed and designed for small local consumers (100kW-1MW) for the application in the geothermal low-temperature sources 150-250C, big ferries (unused hot waters from engine), hot waters and gases from metallurgical and chemical factories, and many other similar customers. The drawings for the construction have been made and the optimal parameters computed for some potential liquid metal working media and two variants of the unit modules' assembling have been elaborated (parallel assembling for getting the desired voltage or consecutive assembling for obtaining the desired current in the electrical network). The optimal height of the liquid metal circulating loop was obtained in the range 10-15 m, and the voltage in a unit 1.2-1.5 Volt.

## 1. Accomplishments and challenges in utilization of low-potential heat sources

The rapid development of modern earthly civilizations requires ever greater resources of energy that humanity is constantly looking for. You need large reserves of clean energy, and it is desirable as cheap as possible. So, recently, for example, the topic of extracting huge reserves of energy and fresh water due to the movement of atmospheric air is rapidly developing in many countries around the globe [l-10]. The Swedish company United Science and Capital (US&C) [9] has recently signed an agreement with the Arizona State University about common promotion of the energy tower proposed by company. Also US&C is now in an advanced stage of negotiations with Governments of many countries with issues in water and food security, as well as for profit and non-profit organizations.

People have been looking for the sources of energy since ancient times when recognized their advantages from using the energy of the sun and fire, later on using the energy of waterfalls, wind and other sources of energy. Viktor Schauberger nearly hundred years ago voted to the mankind with his clear statement to use the implosion instead of explosion, burning the mineral resources of the planet (V. Schauberger: Unsere sinnlose Arbeitet - Die Quelle der Weltkrise, Der Aufbau durch Atomverwandlung, nicht Atomzertrümmerung (1933, Krystall-Verlag GmbH, 2001, Jörg Schauberger, released in English as "Our Senseless Toil - The Cause of the World Crisis - Progress Through Transformation of the Atom - Not its destruction!"). Nowadays there are known many attempts for creation and using the clean energy of the diverse renewable sources instead of burning the coal, oil, etc. destroying and exhausting our beautiful planet.

One of the modern directions in a development and implementation of the new technologies and devices is cogeneration and a use of the low-potential energy sources, which have a low exergy and as such are not suitable for a profitable intensive production of energy. This paper deals with one of such directions connected to utilization of the low-potential energy [11-21], which is presently wasted in a global scale around the world due to the engineering and technical problems of its efficient use. Liquid Metal MagnetoHydroDynamic Energy Converters (LMMHD EC) of gravity type have been recently proposed for a variety of the heat sources for electrical power generation [l1]. In these systems, the vertical two- phase flows consisting of a steam and the high density liquid metals like a lead alloy take place in the riser pipe. The design of an optimum LMMHD EC and a scaling up of the system requires an accurate modeling of the two-phase flows. The most important parameter which governs the two-phase flow is a void fraction. Such approaches for a modeling of the multiphase flows including our works have been developed and proven in the experimental studies [22-35]. The accuracy and the range of applicability of these relate to be verified for liquid metal two-phase flows. The experiments have been conducted in a nitrogen-mercury simulation LMMHD EC facility. The time averaged void fraction was measured for various flow rates using gamma-ray attenuation method. Corrections for the cross-sectional variations and dynamic fluctuations in the void fraction as well as the errors due to finite beam size of the gamma ray have been applied to improve the accuracy of measurement. Measured void fraction and pressure profiles were compared with predicted values based on the well known empirical relations for a void fraction. In addition, the data was compared with the values based on bubble/slug flow models, which take into consideration the momentum equation of the gas explicitly. The results of the analysis have been presented in the paper [13].

The steam in the LMMHD ETGAR (gravitational) system was initially generated by a conventional boiler and then injected through the mixer to the hot lead [12]. In order to further reduce capital and operation costs of the ETGAR systems, it was proposed to inject the thermodynamic fluid in its liquid phase (instead of steam) into the hot liquid metal in the "riser" branch of the loop. The boiling of the volatile liquid in this case occurs under direct contact with the liquid metal, and thus avoids the need for an expensive external steam generator. It is also anticipated that (the boiling process leads to mixing of the two-phase flow and hence a decrease of the slip between vapor bubbles and liquid metal (higher loop efficiency). The applicability of this idea was tested in an integrated experimental system the "OFRA" system described in [12]. It simulated a real LMMHD ETGAR loop and was designed to operate with lead and eutectic lead/bismuth, at temperatures up to 480°C. The following goals of study were stated: direct contact boiling phenomena of a multi-droplet bed of water in lead; Influence of different additives (surfactants, foaming materials etc.), dissolved in the lead, on the two-phase flow characteristics, namely, void fraction distribution and phase velocity ratio, flow configuration and overall system performances; water reaction with molten lead over a wide range of temperatures, pressures and flow rates; lead reaction with the construction materials in the presence of liquid/vapor water. Our project [14] started from these results.

The development of superconducting installations, such as mini-power plants (MPPs), and their justification is a new trend in the energy applications of superconductivity [18], which can enhance the advantages of MPP. For development, construction and testing, the author [18] developed a high-temperature superconducting (HTS) MPP model with a design capacity of 10 kW. Appropriate approaches and tools are needed to promote the industrial application and introduction of HTS power devices on the market, including HTS MPP equipment. Methods of solving problems of automated design effectively help qualitatively and quantitatively evaluate the technical and economic capabilities of HTS devices.

The Qena Paper Industry Company [19] prepared analysis of exergy for various components of a 45 MW cogeneration thermal power plant. Each component was analyzed separately, and exergy losses were identified and quantified for each component of the cogeneration power plant. In addition, the effect of changing the ambient temperature on this analysis was investigated. The results showed that the rate of destruction of exergy increases with increasing ambient temperature in all components of the system. The main contribution to the destruction of exergy occurs from the boiler, and it increases with increasing ambient temperature. With a full load and an ambient temperature of 18 to 45 °C, the percentage of destruction for the boiler, deaerator and condenser was 78.9% to 80.5% and from 3.6% to 3.3% and 3.1% up to 0,6% respectively. The cogeneration power plant is an efficient, clean and reliable approach to the generation of electricity, as well as the supply of thermal energy through a single fuel. In particular, cogeneration, which sometimes uses the waste heat through its processing, achieves a significant improvement in system efficiency. The main advantages of a cogeneration are relatively high efficiency, reducing emissions to the atmosphere, such as nitric oxide ($N_2O$), sulfur dioxide ($SO_2$) and carbon dioxide ($CO_2$). Today, about 80% of the world's electricity is produced approximately from fossil fuels (coal, oil, mazut, natural gas) from thermal power plants, while 20% of electricity is compensated from various sources such as hydraulic, nuclear, wind, solar, geothermal and biogas. Energy consumption is one of the most important indicators, demonstrates the stages of development of countries and the living standards of communities. Population growth, urbanization, industrialization and technological development lead to increased energy consumption, which leads to decisive environmental problems, such as pollution and the greenhouse effect. In recent decades, the exergy index based on the second law of thermodynamics proved to be a useful method in the development, evaluation, optimization and improvement of thermal power plants. Kamate and Gangavati conducted an energy and exergy analysis in a cogeneration plant based at Belgaum in India [19].

ORMAT [17] developed the geothermal recovery mini-power plants based on the organic Rankine cycle for the moderate temperatures (85–150 °C) and some of them were done for the higher temperatures. The power of these MPPs was in a range of 300–1300 kW. These factory-integrated and tested modules of the two or more units can be combined for applications, where the geothermal or industrial waste heat source is economically sufficient to install the larger power plants. Experience has been acquired in operation of the low-enthalpy geothermal power plants in the USA, as well as in other countries. A few typical power plants of the power 800 kW, 3.2 and 30 MW have been presented in [17].

## 2. Liquid Metal Magnetohydrodynamic Gravitational Mini Power Plant

Our paper is devoted to a description of the project and the results obtained as concern to a development of the LMMHD GMPP for utilization of the low-potential heat from any available low-exergy sources, which are the huge sources of the wasted energy around the globe. The project was based on the gravitational vapour/gas-lift driving principle and gained experience by the LMMHD energy transformation with. An example of the GMPP was developed and designed for small local consumers (100kW-1MW) for the application in the geothermal low-temperature sources 150-250 °C, big ferries (unused hot waters from engine), hot waters and gases from metallurgical and chemical factories, and many other similar customers. The drawings for the construction have been made and the optimal parameters computed for some potential liquid metal working media and two variants of the unit modules' assembling have been elaborated (parallel assembling for getting the desired voltage or consecutive assembling for obtaining the desired current in the electrical network). The optimal height of the liquid metal circulating loop was obtained in the range 10-15 m, and the voltage in a unit 1.2-1.5 Volt.

The project was based on the experience by LMMHD energy transformation with gravitational vapour/gas-lift driving principle and the authors' experience in the development of MPP for utilization of low-potential heat performed in Ukraine. The MPP for utilization of low-potential heat was developed with the goal to reduce an energy production cost at the power plants and to utilize the low-temperature heat sources in any available industrial application, e.g. geothermal heat sources, hot waters and gases from metallurgical and chemical factories, big ferries, and so on, where the heat sources with temperature above around 150 $^0$C are available for use.

The structure and functionality of the basic components of MPP can be explained according to the simple schematic representation given in Fig.1. The MPP consists of a circulatory loop made by channels (tubes), two vertical and two horizontal. The diameter of the channels is about 10-20 cm depending on the power and other parameters. The two vertical channels of the circulatory loop are the lifting and falling ones (the left and right one, correspondingly), where the liquid metal (working fluid) is circulating. The bottom horizontal channel of the circulatory loop contains the heater, which accepts the utilized low-potential heat from the heat source (may be liquid or gas, which passes through the heater). The left vertical lifting channel is driving part of the loop. A two-phase (vapour/gas-liquid) flow is going up, while in falling channel the separated pure working fluid is moving down through the Faraday MHD-generator producing the direct current due to electroconductive fluid (liquid metal) passing through the MHD generator. The two vertical channels are connected at the top of an installation with a separator. At the bottom they are connected with a heater utilizing the low-potential heat.

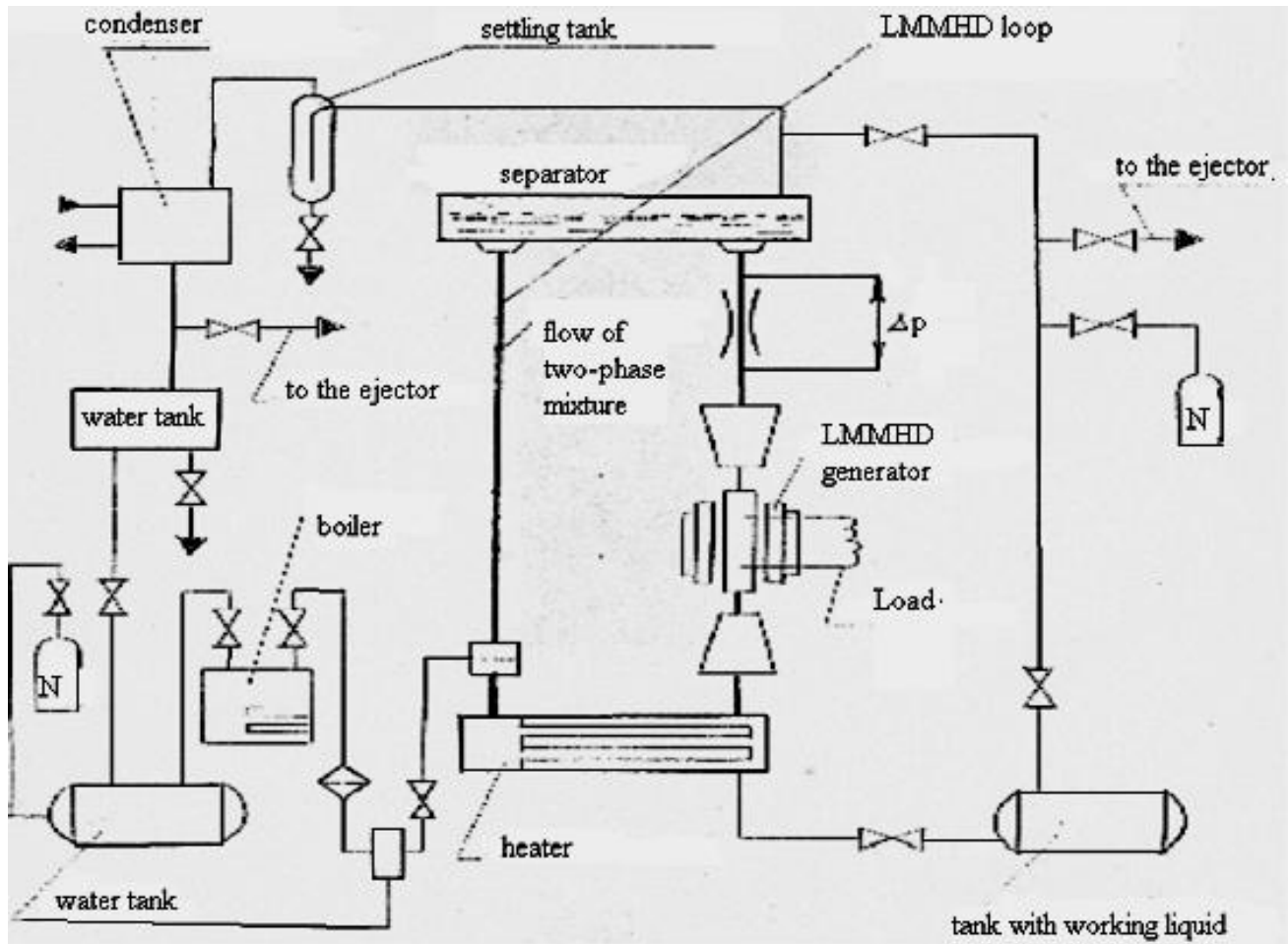

Fig.1. The simplified structural scheme of the MPP.

Circulatory liquid metal (LM) loop is created for a continuous circulation of the working LM due to a difference of the densities in the left and right vertical channels and due to a vaporization of the injected water from the bottom of the left channel. The mixer installed at the bottom of the left channel serves for injection of the thermodynamic fluid (saturated water) into a working fluid in the loop. The heater delivers a heat to the working fluid. Upper part of the channel is connected to a separator where the vapor is separated and supplied to the condensation and cleaning system. Separated working fluid is going to the right vertical channel. Channel of LMMHD-generator contains the confuser, straight forward channel and the diffuser (analogy of the Laval nozzle) for speeding-up the working fluid in the MHD-generator. The Faraday-type induction LMMHD-generator is used, with electric commutation (parallel or consecutive) of the separate autonomous unified modules into electrical energy system. The last one is very useful for practice as far as it allows assembling of the MPP of the desired power and output electric energy from the unified modules.

The facility can work autonomously as well as together with another similar devices having consecutive or parallel electrical commutation of their MHD-generators. The electric facility can be equipped with the transformer of a direct current to an alternating current. The estimated parameters of MPP are as follows:

- The initial temperature of a heating fluid (water, gas, vapour) in a circulatory loop is 120-180 °C
- The upper temperature of the cycle – 150-250 °C
- Lower temperature of the cycle – 65 °C
- Vapor pressure – 5 bar
- Power of the Plant – 0.1-1.0 MW
- Height of the channels, 7-20 m
- Distance between the channels of MPP, more than 0.5 m

By preliminary estimations from our construction and calculation work, the LMMHD-generator has to have the following parameters:

- Length of the electrodes – 0.5 m
- The magnetic field – 0.1-1.0 T
- The voltage – 50-100 V (can be regulated by a stated number of the assembled unified modules and the type of an electric commutation applied: parallel or consecutive)
- The capacity, current and voltage are regulated by a desire of the consumer under assembling of the MPP through a number of the assembled unified modules and the type of their commutation chosen.

## 3. Modeling of Thermal Hydraulic Processes in MPP

The project aimed on increasing the energy efficiency through utilization of the low-potential heat sources, which are presently in a wide amount not in use or are used partially and in an inefficient way. The MPP was be developed and designed for small local consumers (100kW-1MW) in the areas, where any low-potential heat sources are available, for example, geothermal hot water with temperatures about 120-180 $^0$C, big ferries (unused hot waters from engine), hot waters and gases from metallurgical and chemical factories, etc. First of all, the investigation of the processes was performed, with the calculation and optimization of the MPP' parameters, e.g. geometry (the height and diameter of the channels, size of the circulatory loop, the thermal hydraulic parameters of a liquid metal flow in the loop, heat transfer processes in the heater and in the loop, vaporization of injected water, etc. The parameters of the LMMHD-generator were computer modeled and optimized. Altogether parameters of the MPP have been simulated and optimized afterwards.

One of the most important sections of MPP is the driving vapor/gas-lift flow in the left vertical channel. Preliminary investigation has shown that this part may play a key role as the MPP's efficiency is changing depending on the evaporation regime and the liquid-bubble lifting flow regime. Injected water is heated in the hot liquid metal inside the channel and evaporates. This leads to a vapor expansion and a vapor lifting force in the left channel, which strongly depends on the size and distribution of the vapor bubbles in the flow. In different conditions the efficiency of MPP may change dramatically, e.g. ten times and more. To optimize investigation of the processes I a heterogeneous two-phase flow of the liquid metal with vapor bubbles the methods developed and elaborated in [22, 28-32] seem to be the most suitable. According to the method of heterogeneous multiphase flows developed by Prof. A.I. Nakorchevski [22] the characteristics of a mixture $a^l(t)$ (mass, velocity, impulse, etc.) of the corresponding characteristics of different phases $a_i^l(t)$ in a multiphase flow are expressed in next form

$$a^l(t) = \sum_{i=1}^{m} B_i(t) a_i^l(t),$$ (1)

where $B_i(t)$ are co-called function-indicators of the phases in multiphase mixture determined as

$$B_i(t) = \begin{cases} 1, if & i-\text{phase occupies the elementary volume } \delta V \\ 0, if & i-\text{phase is outside the elementary volume } \delta V \end{cases}. \qquad (2)$$

With this approach, the analog of the Navier-Stokes equations in a boundary layer approximation was derived:

$$\frac{\partial}{\partial x}(y\rho_i B_i u_i) + \frac{\partial}{\partial y}(y\rho_i B_i v_i) = 0, \qquad \sum_{i=1}^{m} B_i = 1, \qquad \rho_i B_i (u_i \frac{\partial u_i}{\partial x} + v_i \frac{\partial u_i}{\partial y}) = -\frac{dp}{dx} + \frac{1}{y}\frac{\partial}{\partial y}[yB_i\tau_i]_m, \quad (3)$$

where the sum by mute index $i$ is taken by all phases $i$. In the stationary equations of incompressible liquids (3) written in a cylindrical coordinate system are: $p$- pressure, $\rho$- density, $u,v$- the longitudinal and transversal velocity components, $\tau_i$- turbulent stress for a phase $i$. Index $m$ belongs to the values at the axis of the flow (symmetry axis). All values are averaged on the given interval by time. The schematic representation of the flow of two immiscible liquids is given in Fig. 2 for example of turbulent two-phase jet of two immiscible liquids:

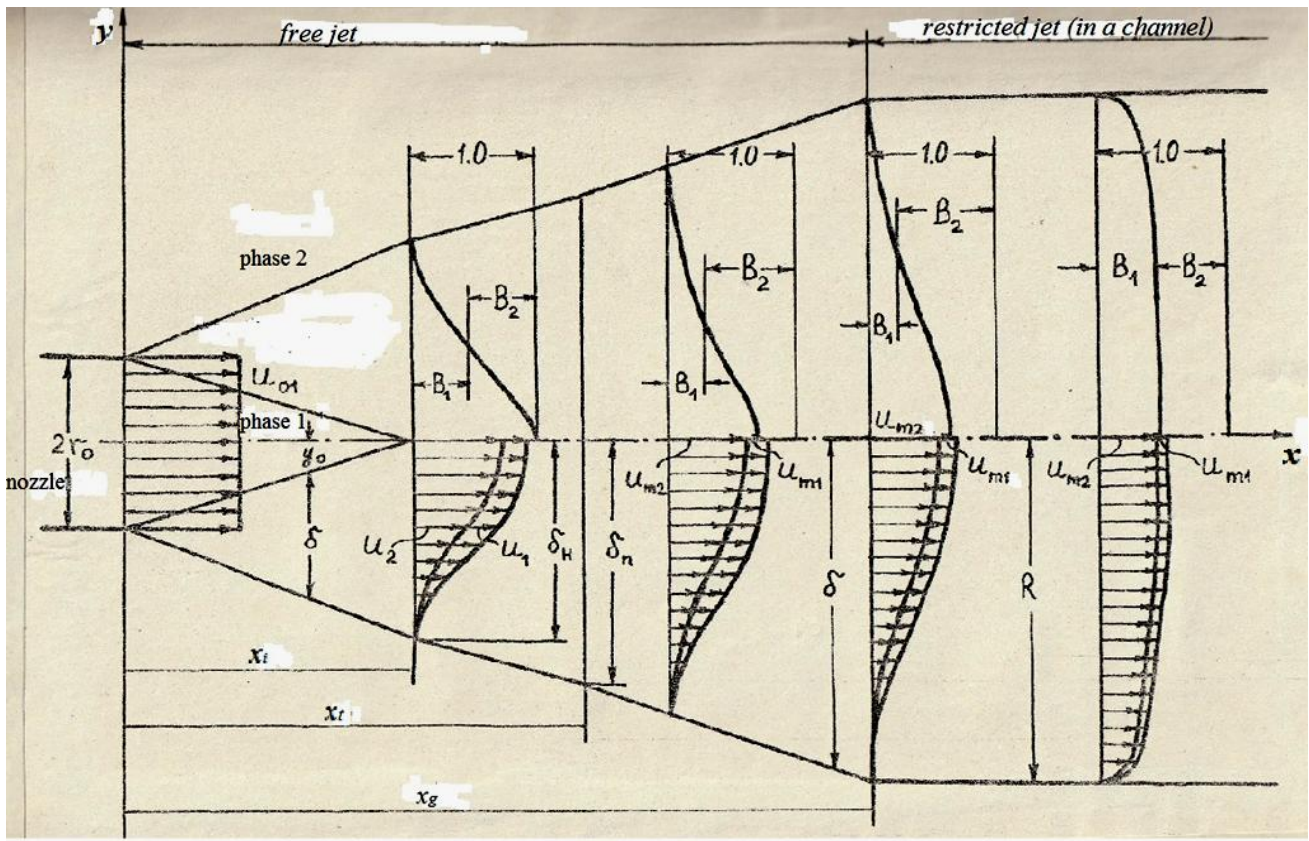

Fig. 2 The structural scheme of the two-phase jet of immiscible liquids

One liquid is going from the nozzle of the radius $r_0$ with velocity $u_{01}$ (the velocity profile is supposed simple uniform) in a surroundings occupied by other liquid (phase 2) being in a rest. The structure of a jet's mixing with surrounding liquid according to the Fig. 2 is simplified according to a traditional scheme [23]. First, the initial part of the length $x_i$ with the approximately linear boundaries for the conical surface (in cylindrical coordinates) of the internal potential core of a first phase and the external interface (conical too) are considered as the boundaries of the mixing area and the central potential core, correspondingly. The mixing turbulent zone between the above surfaces contains drops and fragments of the phases as far as immiscible liquids have behaviors like the separate phases, with their interfacial multiple surfaces interacting in all such locations (exchange of mass, impulse, energy among the phases).

Close to the external interface there is a flow of a surrounding liquid with some drops of the first phase, while, in turn, close to the boundary of the potential core there is a flow of the first phase with the drops of a second phase. As far as we have mutually immiscible liquids, the mixing zone contains two phase mixture in a turbulent flow. After an initial part of the mixing zone when all first phase in a potential core is spent, the short transit area is preparing the ground part of the turbulent two-phase jet, where both phases are well mixed across all layer. Normally for description of multiphase flows the spatial averaging of the differential equations of mass, impulse and energy conservation is performed using the concept of volumetric phase content [24-26], which does not fit so well to experimental study of the separate phase movement in a mixture as the approach proposed in [22], where the special experimental technology and micro sensor for a measurements in two-phase flows has been developed as well. In [25, 26] fundamentals and analysis of the different methods for modeling of the multiphase systems can be found. Actually the known methods of multiphase methods are well connected including the one in [22], and the parameters averaged by time can be easily transformed to the ones by [24-27].

The external interface of the mixing zone is determined zero longitudinal velocity of the second phase and transversal velocity of the first phase (the second phase is sucked from immovable surrounding into the mixing zone). The function-indicator of the first phase $B_1(t)$ is zero at the external interface because the first phase is absent in surroundings. Similar, the function-indicator $B_2(t)$ is zero on the interface of the potential core, boundary of the first phase going from the nozzle. In a first approach, an influence of the mass, viscous and capillary forces is neglected. With account of the above-mentioned, the boundary conditions are stated as follows [22]:

$$y=y_0, \ u_i=u_{0i}, \ v_i=0, \ \tau_i=0, \ B_1=1; \qquad y=y_0+\delta, \ u_i=0, \ v_i=0, \ \tau_i=0, \ B_1=0. \qquad (4)$$

The turbulent stress in the phase is stated by the "new" Prandtl's formula

$$\tau_i=\rho_i \kappa_i \ \delta u_{mi} \partial u_i \ / \ \partial y \ , \qquad (5)$$

where $\kappa_i$ is the coefficient of turbulent mixing for $i$-th phase, $\delta$ is the width of the mixing layer.

The polynomial approximations for the velocity profiles and other functions in the turbulent mixing zone have been obtained based on the boundary conditions (4) in the form:

$$u_1 / u_{01} = 1 - 4\eta^3 + 3\eta^4, \qquad u_2 / u_{02} = 1 - 6\eta^2 + 8\eta^3 - 3\eta^4, \qquad (6)$$

$$B_1 = B_1^{(0)} = 1 - \eta^3 + 0.5\eta^2(1-\eta)h(x), \qquad h \in \left[-6, 0\right],$$

$$B_1 = B_1^{(1)} = 1 - 4\eta^3 + 3\eta^4 + 0.5\eta^2(1-\eta)^2 h(x), \qquad h \in \left[-12, -6\right],$$

$$B_1 = B_1^{(2)} = 1 - 10\eta^3 + 15\eta^4 - 6\eta^5 + 0.5\eta^2(1-\eta)^3 h(x), \qquad h \in \left[-20, -12\right], \qquad (7)$$

$$B_1 = B_1^{(3)} = 1 - 20\eta^3 + 45\eta^4 - 36\eta^5 + 10\eta^6 + 0.5\eta^2(1-\eta)^4 h(x), \qquad h \in \left[-30, -20\right],$$

$$B_1 = B_1^{(4)} = 1 - 35\eta^3 + 105\eta^4 - 126\eta^5 + 70\eta^6 - 15\eta^7 + 0.5\eta^2(1-\eta)^5 h(x), \qquad h \in \left[-42, -30\right],$$

$$B_1^{(5)} = 1 - 56\eta^3 + 210\eta^4 - 336\eta^5 + 280\eta^6 - 120\eta^7 + 21\eta^8 + 0.5\eta^2(1-\eta)^6 h(x), \qquad h \in \left[-56, -42\right],$$

$$B_1^{(6)} = 1 - 84\eta^3 + 378\eta^4 - 756\eta^5 + 840\eta^6 - 540\eta^7 + 189\eta^8 - 29\eta^9 + 0.5\eta^2(1-\eta)^7 h(x), \ h \in \left[-72, -56\right],$$

where $h(x) = \left( \partial^2 B_1 / \partial \eta^2 \right)_{\eta=0}$ is an interesting function, which determines a transition of the piecewise continuous function-indicator $B_1^{(n)}$ to its next approximation, determined from the condition that the derivative by $\eta$ with respect to a point $\eta = 1$ be equal to zero up to $(n+1)$-th and including order. Based on these approximations the integral correlations have been derived for the two-phase turbulent jet [22]:

$$u_{01}\left(r_0^2 - y_0^2\right) = 2\delta \int_0^1 B_1 u_1 \left(y_0 + \delta\eta\right) d\eta, \quad \rho_1 u_{01}^2 \left(r_0^2 - y_0^2\right) = 2\delta \int_0^1 \left(\rho_1 B_1 u_1^2 + \rho_2 B_2 u_2^2\right)\left(y_0 + \delta\eta\right) d\eta,$$

$$\rho_1 u_{01}\left(u_{01} - u_1^*\right) y_0 y_0^{'} + \frac{d}{dx}\delta \int_0^{\eta^*} \sum_{j=1}^2 \rho_j B_j u_j^2 \left(y_0 + \delta\eta\right) d\eta - \sum_{j=1}^2 u_i^* \frac{d}{dx}\delta \int_0^{\eta^*} \rho_1 B_i u_i \left(y_0 + \delta\eta\right) d\eta =$$

$$= \left(y_0 + \delta\eta^*\right)\sum_{j=1}^2 \rho_j B_j \kappa_j u_{0j} \frac{\partial u_j^*}{\partial \eta}, \qquad B_1 + B_2 = 1. \qquad (8)$$

The first equation in (8) was got integrating by $y$ the mass conservation equation, the second and the third ones – integrating the impulse conservation for the total flow of a two-phase mixture for $y=y_0+\delta$ and $y=y^*$, respectively. The polynomial approximations for the functions $u_2$, $B_1$ on a ground part of the jet keep the same but for the function $u_1$ approximation and the integral correlations for the ground part of a jet are [22]:

$$u_1 / u_{m1} = 1 - 3\eta^2 + 2\eta^3, \quad 2\int_0^\delta B_1 u_1 y dy = u_{01} r_0^2, \quad 2\sum_{j=1}^2 \int_0^\delta \rho_j B_j u_j^2 y dy = \rho_1 u_{01}^2 r_0^2, \qquad (9)$$

where the first is equation of the mass conservation for the first phase, the second and the third – the momentum conservation equations for the total and for the part of the cross section, respectively, according to the methodology [23]. And the momentum equation on the jet's axis ($y=0$) is used too:

$$\sum_{j=1}^2 \rho_j B_{mj} u_{mj} \frac{du_{mj}}{dx} = 2\sum_{j=1}^2 \left[ \frac{\partial}{\partial y}\left(B_j \tau_j\right) \right]_m. \qquad (10)$$

The mathematical model including the ordinary differential equations (8)-(10) by longitudinal coordinate $x$ are implemented for analysis and numerical simulation on computer the basic features of the stationary turbulent two-phase jet of two immiscible liquids. The function-indicator $B_1$ shows how much is a presence of the first phase in a selected point of mixing zone, which can be directly compared to an experimental data by two-phase sensor. Therefore a solution of the task may give both parameters of the flow together with their belonging to a particular phase. Examples of computer modeling and comparison of the results obtained against the experimental data for turbulent heterogeneous jets are given in Fig. 3 and Fig. 4 for different density ratio of the mixing liquids and available slip of phases.

As shows the Fig. 3 this method reveals good correlation with experimental data. The intensity of the phase mixing is seen from the Fig. 4. The density ratio 8.0 can approximately correspond to vapor/liquid melt of the MPP accounting for available slip of the phases. After optimization of thermal hydraulic parameters, the geometry of MPP was done and then the parameters of the LMMHD-generator have been determined. This served as a basis for a design of the construction of MPP according to the specified needs of consumer. The construction was performed in such a way that the MPP was proposed to be always assembled from the unified modules, which allows assembling the required plant with account of the local consumer's needs in each specific case. Thus, the MPP supposed to be mobile-based and easily transported to a location of the consumer and then assembled locally [36].

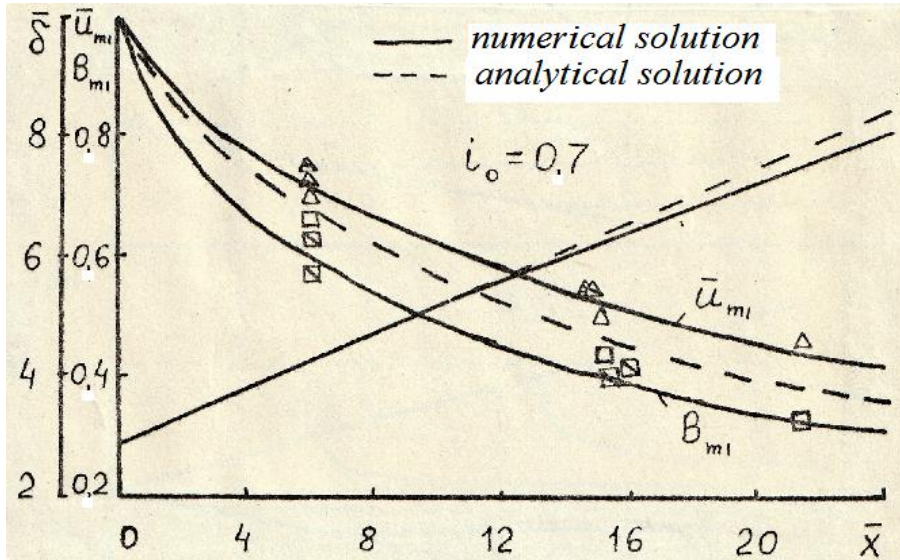

Fig. 3 Parameters of turbulent two-phase jet against experimental data for oil-water immiscible liquids

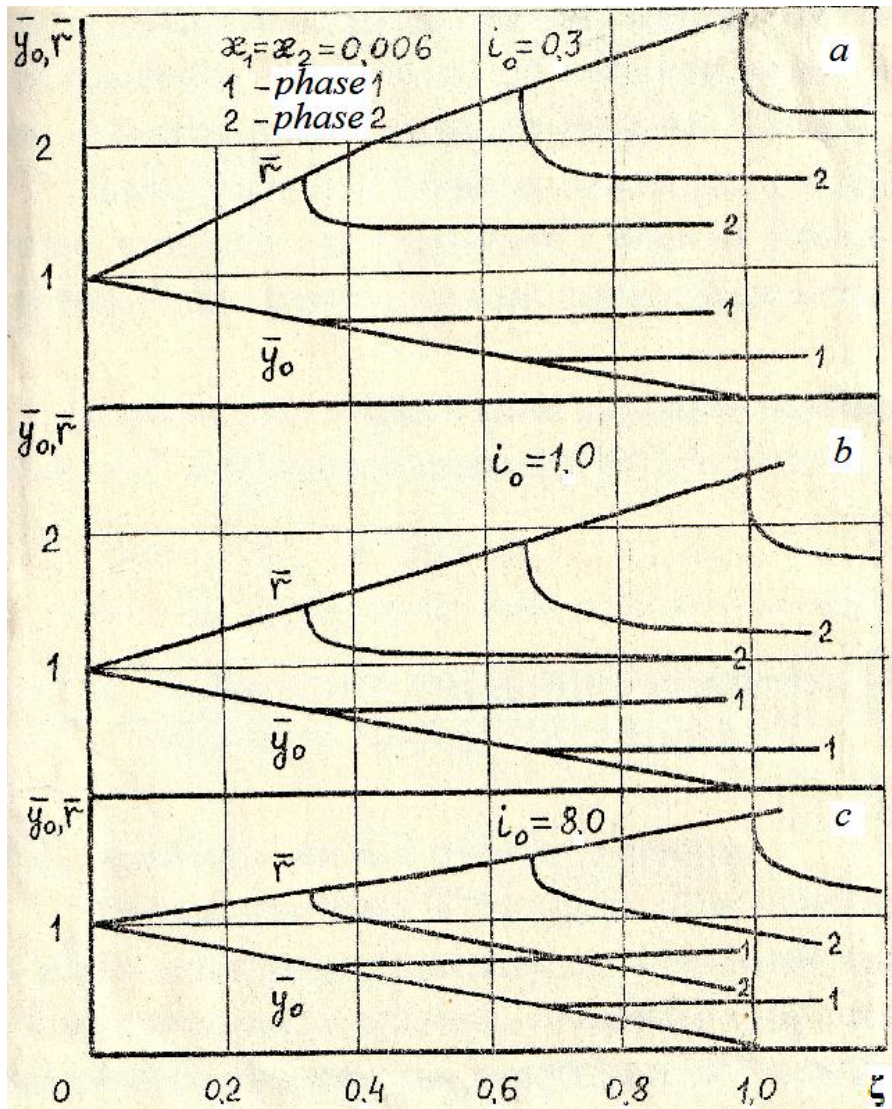

Fig. 4 The stream lines in a turbulent two-phase jet of the immiscible liquids

The research was carried out through the fundamental analysis and the numerical experiments in a wide range of the parameters of interest. The design has been done in a way to achieve the movable MPP construction, which is easily assembled from the unified modules in any required place according to the needs of consumer and specific low-potential heat source available there for utilization to produce electricity. The cost for the project was estimated at $1 mln. Each of the 8 stages of the project was estimated approximately at 3 months duration. Cost estimate was made after discussion of the proposal with the contractor based on the specific task stated.

## 4. The exergy and the new molecular thermodynamic theory by Eroshenko V.A.

With regard to the exergy of heat available at a given temperature, we must indicate that the maximum possible conversion of heat to work or the content of exergy, depending on the temperature, depends on the temperature at which the heat is and the temperature at which heat can be used, the latter is ambient temperature. Carnot (1824) found the upper limit for the transformation of heat into work [37], which is known as Carnot's efficiency: $\eta = 1 - T_C / T_H$ (absolute temperatures), where $T_H$ is a higher temperature, and $T_C$ is the lower temperature. Exergy exchange: $B = Q(1 - T_0 / T_s)$, where $T_s$ is the temperature of the heat source, $T_0$ is the ambient temperature. It is obvious that the cycle's efficiency is maximized by the maximum of $T_H$ and the minimum $T_C$. In the 1870s, one of the most prominent scientists, J.W. Gibbs (1873) united Thermochemistry into a single theory [38], incorporating a new concept of chemical potential to evoke a shift away from chemical equilibrium into Carnot's work in describing thermal and mechanical equilibrium and their potential for change. The unified Gibbs theory has led to the functions of the state of thermodynamic potential, which describes the differences from the thermodynamic equilibrium. Gibbs received the mathematics of the available energy of the heat source in the form above. Since then, the physics that described exergy has not changed much.

The actual amount of available energy depends on conditions that cannot be simulated. But for a real system, energy is saved, and exergy is energy available for useful work. As we mentioned above, there is a lot of energy around the world (solar, wind, hydropower, geothermal, industrial heat loss, etc.), but for various reasons it is inaccessible, just some of it. Thus, energy consists of two parts: exergy (an available part of energy) plus anergy (inaccessible part). For more details on the concepts of energy and exergy and the differences about them, shown in several cases considered with calculations, there are, for example, references to present the concept of exergy and its applications [39, 40]. Understanding the concept of exergy and its potential is very useful for conducting an exergy analysis of the real systems or processes.

One of the most exciting theories on the end of XXth century is molecular thermodynamics founded by Professor of the NTUU "KPI" Valentyn Andriyovych Eroshenko [41-49] who revealed the huge potential of a surface energy as a new working medium instead of used for the centuries volumetric energy of a working bodies. Gas/vapor has been used as a working body of all heat engines, turbines, heat and power facilities, mechanical power batteries and other devices and thermo mechanical systems for two centuries in many branches of industry (energy, engineering, space and defense industry, etc.).

All thermodynamic transformations are based on kinetic energy of a chaotic thermal movement of gas/vapor molecules. Intermolecular potential energy is zero for an ideal gas, and negligible for real gases. For the first time in the world it was proposed to use the **Interfacial surfaces** in heterogeneous lyophobic systems as the **new working bodies** in the thermo-mechanical devices and systems for energy accumulation, dissipation and conversion. In the new working bodies named **repulsive clathrates (RK)**, the potential energy of Intermolecular interaction becomes predominant, while the kinetic energy of the molecules is less interesting as the lower one. Because the forces and Intermolecular energy potential depend on the temperature, the new direction of technical thermodynamics has been called "Thermomolecular energy".

In terms of the technology repulsive clathrates are considered the condensed systems "liquid + capillary-porous matrix unmoisten to this liquid" (contact angle $\theta \gg 90$ deg). Extensive parameter of the system is interfacial surface $\Omega$, and intensive parameter - the surface tension of the liquid $\sigma$. The matrices are characterized by specific surface (from 200 to 1000 m$^2$/g) and by radiuses of the capillaries and pores r (from 0.3 to 10 nm). As the porous matrices Silica gels, Silchromes Al$_2$O$_3$, porous glasses, zeolites and other carriers of a well developed specific surface are used. Porosity of the matrices applied is from 0.2 to 0.7 cm$^3$/cm$^3$. The fluids used are as follows: water and aqueous solutions, electrolytes, salts, low temperature eutectics and metal alloys. This new direction in the thermodynamics is the most promising in the reveling the huge potential of the surface energy and low-exergy sources of energy totally.

All thermodynamic transformations are based on the thermomechanical processes of the interfacial surface $\Omega$ "extension-reduction" with the work $\delta W$ and heat $\delta q$: $\delta W = \sigma \cos\theta \, d\Omega$ - the work by isothermal surface development (the Gibbs work); $\delta q = T \cdot (d\sigma/dT) \cdot \cos\theta \, d\Omega$ - heat of the isothermal surface ($T$- temperature of a process, $d\sigma/dT$ is the temperature coefficient of a surface tension of a fluid, the physical constant). The specifics of the RK are that the pressure is a subject to the hydraulic Pascal law and is defined by the hydraulic capillary Laplace pressure. Changing the volume of the condensed heterogeneous system $\Delta V$ is determined by the rate of filling the pore space, which is the rate of a development of the interfacial surface $\Omega$, and it is not determined by a deformation process. The total benefit from the creation and use of thermomolecular devices, for example motors invented by V.A. Eroshenko should be connected to a radical saving of the fuel and construction materials, as well as to a reducing the heat and chemical pollution: the thermomolecular engine (TME) works from the external heat supply. This makes the TME similar to the Stirling engine, which one with the humankind's last hope is connected as for the solution of the environmental problems: "Stirling will save the world!". But the three versions of the TME patented by Eroshenko are substantially more effective and they are available for producing nowadays.

## 5. The conclusions

The Liquid Metal Magnetohydrodynamic (LMMHD) Gravitational Mini Power Plant (GMPP) for utilization of the low-potential heat from any available low-exergy sources presented in the paper may be recommended for practical implementation after its technological finalization for producers of the ones. The main its advantage is possibility for utilization of any sources of the low/potential heat and mobile module principle of the assembling. It may be applied around the world for use of the geothermal low-temperature sources (150-250 Celsius degree), big ferries (unused hot waters from engine), hot waters and gases from metallurgical and chemical factories, etc. The optimal liquid metal applied as a working media in the MPP requires some more investigation for optimization from a few points of view. Two variants of assembling of the unit modules are proposed for the customer (parallel assembling for getting the desired voltage or consecutive assembling for obtaining the desired current in the electrical network). The optimal height of the liquid metal circulating loop was obtained in the range 10-15 m, and the voltage in a unit 1.2-1.5 Volt (the total voltage can be got any desired by consecutive assembling of the units).